\documentstyle[11pt,aaspp4,here]{article}

\received{03 November1998}
\accepted{29 January 1999}
\slugcomment{To appear in May 1999 issue of {\it The Astronomical Journal}} 

\lefthead{Ratnatunga, Griffiths, Ostrander }
\righthead{Gravitational lenses from HST MDS}

\def\M  {{$-$}}
\def\VT  {{$V_3$}}
\def\degpt {{$\buildrel{\circ}  \over .$}}

\begin{document}

\title {The Top Ten List of Gravitational Lens Candidates \\
from the HST Medium Deep Survey}

\author{Kavan U. Ratnatunga, Richard E. Griffiths, Eric J. Ostrander}

\affil{Physics Dept., Carnegie Mellon University, \\
Pittsburgh, PA 15213 \\
 kavan, griffith, \& ejo@astro.phys.cmu.edu}

\begin{abstract}
A total of 10 good candidates for gravitational lensing have been
discovered in the WFPC2 images from the HST Medium Deep Survey (MDS) and
archival primary observations. These candidate lenses are unique HST
discoveries, i.e. they are faint systems with sub-arcsecond separations
between the lensing objects and the lensed source images. Most of them
are difficult objects for ground-based spectroscopic confirmation or for
measurement of the lens and source redshifts. Seven are ``strong lens''
candidates which appear to have multiple images of the
source. Three are cases where the single image of the source galaxy has
been significantly distorted into an arc. The first two quadruply lensed
candidates were reported in \cite{rat95}.  We report on the subsequent
eight candidates and describe them with simple models based on the
assumption of singular isothermal potentials. Residuals from the simple
models for some of the candidates indicate that a more complex model for
the potential will probably be required to explain the full structural
detail of the observations once they are confirmed to be lenses. We also
discuss the effective survey area which was searched for these
candidate lens objects.
\end{abstract}

\keywords {cosmology:observations - gravitational lensing - surveys}
 
\section{Introduction}

The HST Medium Deep Survey (MDS) (\cite{gra94,grb94,rat99}), has
comprised parallel WFPC2 observations of just over 400 random fields
for the systematic study of the evolution of faint galaxies, as well
as being a serendipitous survey which has resulted in the discovery of
many interesting objects. The survey has provided a unique set of data
for over 150,000 galaxies wherein the selection of a correspondingly
large number of field galaxies has been possible down to I$<$25
(\cite{gri96}).  The discovery of two quadruple-type lenses from the
HST MDS, viz.  HST 12531$-$2914 and HST 14176$+$5226 has previously
been reported (\cite{rat95}) and subsequent spectroscopic observations
of HST 14176$+$5226 have provided confirmation that the system is
indeed a gravitational lens (\cite{cra96,mou98}).

Observations and modeling of lens systems can be used to constrain
directly the distribution of dark matter in the lensing galaxies. The
characteristic separation $\Delta\theta$ of lensed images depends on
the total lensing mass and the distances between the observer, the
lens, and the source object. This allows the mass-to-light ratio of
the lens galaxy to be determined directly if the lens and source
redshifts are known. The shape of the lensing mass distribution can
also be constrained, using the configuration and light distribution of
the lensed images.

Gravitational lenses can also be used to provide constraints on the
geometry of the universe. The image separations $\Delta\theta$ depend
on the angular size distances to the lens and background source, which
in turn depend on global cosmology. In order to measure the
cosmological constant $\Lambda$, for instance, it has been suggested
that strong gravitational lenses might be used, i.e. isolated galaxies
or clusters of galaxies for which the gravitational potential results
in multiple imaging of a background source object
(\cite{pac81,alo86,got89}). The use of the lens number counts (or the
optical depth) has also been advocated since this is very sensitive to
$\Lambda$ (\cite{fuk92}).  \cite{mao93,koc96} have applied this
method, obtaining upper limits of $\Lambda \lesssim 0.7$. Predictions
of lensing probabilities by galaxies are very sensitive to
$\Lambda$. If $\Lambda\simeq1$, for example, we would expect to see
about 10 times more lensed systems than if $\Lambda=0$
(\cite{mao93,koc93}).

Lensing constraints on cosmology are not entirely free from systematic
errors, however. The biggest problem is the lack of information on the
lens galaxies at $z > 0.4$, which still allows significant error in
the estimate of $\Lambda$. Elliptical galaxies dominate the expected
lensing rates, due to the large concentration of mass near their
centers (e.g. \cite{kee97}), so any statistical bias needs be
determined by comparing the non-lensing elliptical galaxies at
comparable redshifts.

At the faint magnitudes $V>20$ surveyed by the HST MDS we might expect
to discover a few gravitational lenses in each square degree of
sky. Unfortunately the total area of sky surveyed by the WFPC2 is
still under about 1 square degree, and most of the lens candidates
discovered are too small and faint for ground based spectroscopic
follow-up with even the largest telescopes. The sample we are locating
today probably needs the next generation of space based
instrumentation for more detailed studies.

\section{Gravitational Lenses in the Medium Deep Survey}

\cite{mir92} have pointed out that, for the same lens galaxy
population that is responsible for the known lensed radio sources,
there should be $\sim100$ lensed faint galaxies per square
degree. With small image separations ($\Delta\theta\simeq1\farcs$) and
faint background sources, these lensed systems cannot be detected from
the ground. However, they should be readily detectable using HST
observations, and could represent the largest untapped mine of lensed
systems. A large statistical sample would greatly increase the
importance of lensing for studies of galaxy evolution and cosmology.

Among the sample of field galaxies in the MDS, we have discovered ten
gravitational lens candidates. The discovery of two lenses HST
12531$-$2914 and HST 14176$+$5226 (Figures~1-2) has been reported
already (\cite{rat95}) and the spectroscopic observation of HST
14176$+$5226 has provided confirmation that the system is a
gravitational lens (\cite{cra96,mou98}). The other eight lens
candidates were discovered subsequently, and their HST images are
shown in Figures~3-10, in the light of filters F814W (I), F606W (V)
and, when available, F450W (B). Each image extracted from an HST WFPC2
observation is $6\farcs4$ on each side. Typically, the lensed (source)
images are blue arcs, while the lensing galaxy appears to be a red
elliptical galaxy, thus making them very good candidates for
gravitational lensing.

\section{Simple models of the Gravitational Lens systems}

A two-dimensional image fitting procedure was developed to model the
first two HST lenses, as described in \cite{rat95}. This simple
modeling approach used singular isothermal ellipsoid potentials
(\cite{kor94}), and this same approach was applied to each of the
gravitational lenses to derive the best fitting model in each case.
Our approach does not require careful astrometry or photometric
measurement of the faint, under-sampled WFPC2 images. The model is
derived by fitting the full 2D-image rather than a list of estimated
positions and magnitudes. This approach, which is more CPU intensive,
makes use of the full extended nature of the light distribution of
lens and lensed images and also uses the observational errors
associated with each image pixel. The choice of the number of
parameters fitted was done interactively so as to obtain a good
convergence while not exceeding the number of significant parameters.

In a few cases the system was observed in several filters and the
lensed images in through some filters were too faint or not
sufficiently well resolved from the brighter lens galaxies. For these
cases it was not possible to fit all the parameters and for the
adopted model we took the geometric parameters and the description of
the potential from the filter in which the system was observed best.

The centroid of the gravitational potential of the lens was assumed to
be the same as that of the light of the lensing galaxy. The axis ratio
and orientation of the mass (potential) were left as free parameters
in the model. If the lensed object is extended with a half-light
radius larger than a pixel, then the axis ratio and orientation of the
lensed galaxy were left as free parameters. The image profiles of the
lens and the lensed objects were selected to be Disk-like or
Bulge-like. If the lensed object was point-like within the WFPC2
resolution then a Gaussian profile was adopted. In a few cases the
lens was better fitted by a disk-plus-bulge model as could be seen
when the lens galaxy was analyzed alone through the regular MDS
pipeline analysis software. However, to keep the modeling simple and
limit the number of parameters fitted, the lens was modeled with a
single component.

Images of the candidate lenses in each filter are illustrated for
F814W (I), F606W (V) and when available F450W (B), from the top
down. On each horizontal strip of five objects, we have from left to
right 1) the region selected for analysis, 2) the model of the WFPC2
observation, 3) the model without convolution by the HST PSF, 4) the
residual image after subtraction of the model image from the observed
image and 5) the model without the lensing galaxy.
% Some of these can be omitted for publication. Maybe only 1, 2 and 4
The corresponding tables give the fitted parameters, with errors for
each model. More details and FITS files of the observations and fitted
models are available from the MDS website http://archive.stsci.edu/mds/.

\section{Discussion of each lens system}

%HST 14176+5226 14:17:36.3 +52:26:44 19.7      u26x8:0009  Cross         ejo      Done Strong 0.083 0.092 Confirmed

\begin{figure}[H]
%\plotone{u26x8g0009.eps}
\plotone{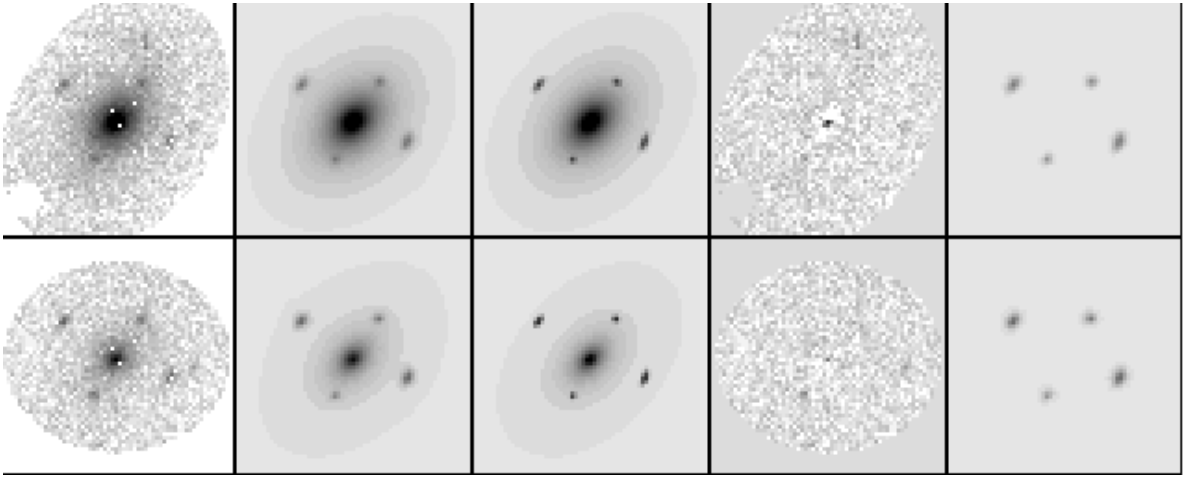}
\medskip
\caption{HST 14176+5226,
The first HST lens candidate in the GWS which has
now been confirmed spectroscopically.
The lens:  F814W=19.8, F606W=21.7 ;
The cross: F814W=24.8, F606W=25.4 }
\end{figure}

{\bf HST 14176+5226} is the first MDS HST ``Einstein Cross'' lens
candidate, found in the ``Groth-Westphal strip'' (GWS). This is the
brightest lens system discovered with the HST and it has been
confirmed spectroscopically. The elliptical lensing galaxy has a
redshift of z=0.803 and a color index (V-I)=1.9 mag. The lensed source
at redshift z=3.4 (\cite{cra96,mou98}) appears to be a QSO with an
apparent color (V-I)=0.5 mag. The impact radius is 0\farcs13 (angular
distance between source and centroid of lens). The image magnification
is about 2.4 magnitudes, making each component about 0.9 mag brighter
than the source QSO. The modeled image gives a very good fit to the
observed image, with a normalized $\chi^2$ of unity.

%\clearpage
%\begin{tabular}{l r r c c c c r r c c c c }
%Configuration &  X     & Y    &   V   & $\pm$ & V\M I  & $\pm$ \\
%Elliptical    &   0    &  0   & 21.68 & 0.04  & 1.97   & 0.04  \\
%   A          & \M 12  & 10   & 25.63 & 0.06  & 0.51   & 0.10  \\
%   B          &  17    & \M 5 & 25.77 & 0.07  & 0.39   & 0.11  \\
%   C          &   9    & 11   & 25.99 & 0.08  & 0.52   & 0.13  \\
%   D          &  \M 4  & \M 10& 25.97 & 0.08  & 0.42   & 0.14  \\
\begin{center}
\begin{table}[H]
\caption{}
\medskip
\begin{tabular}{ l c l c l }
\tableline
Name & \multicolumn{4}{c}{HST14176+5226}\\
\tableline
\tableline
Equatorial(J2000) & \multicolumn{1}{c}{14:17:36.3} & \multicolumn{1}{c}{+52:26:44} & \multicolumn{2}{c}{\VT=32\degpt93}\\
HST WFPC2 &\multicolumn{4}{c}{Groth GTO:5090 11-Mar-1994}\\
Dataset[g][x,y] & \multicolumn{4}{c}{U26X0801T[3][242,700]}\\
MDS Field:id lens galaxy & \multicolumn{4}{c}{u26x8:0009}\\
\\
HST WFPC2 Filter              &\multicolumn{2}{c}{F814W}   &\multicolumn{2}{c}{F606W}   \\
Exposure seconds              &\multicolumn{2}{c}{4 x 1100}&\multicolumn{2}{c}{4 x  700}\\
% log (integrated S/N Ratio)  &\multicolumn{2}{c}{3.57}    &\multicolumn{2}{c}{3.10}    \\
 Fitted Parameter             &  MLE         & $\pm$ rms   &  MLE         & $\pm$ rms    \\
\\
% Sky Mag. ($arcsec^{2}$)     &  22.236      & 0.003       &  22.775      & 0.003       \\
 Total Mag of lens galaxy     &   19.80      &  0.01       &   21.74      &  0.03       \\
% X centroid 0\farcs1 Pix     &  242.92      &  0.02       &  242.79      &  0.03       \\
% Y centroid 0\farcs1 Pix     &  699.92      &  0.02       &  700.09      &  0.04       \\
 Half-light radius            & 1\farcs004   & 0\farcs017  & 1\farcs184   & 0\farcs032  \\
 Orientation of light         & \M 41\degpt5 &   1\degpt0  & \M 41\degpt9 &   2\degpt6  \\
 Axis Ratio of the light      &    0.69      &  0.01       &    0.65      &  0.03       \\
 Mag. of lensed source        &   25.72      &  0.08       &   26.19      &  0.06       \\
 Source X offset 0\farcs1 Pix &   \M 0.28    &  0.05       &   \M 0.19    &  0.05       \\
 Source Y offset 0\farcs1 Pix &    1.20      &  0.05       &    1.36      &  0.04       \\
 Critical Radius              & 1\farcs489   & 0\farcs006  & 1\farcs496   & 0\farcs004  \\
 Axis Ratio Lens mass         &    0.40      &  0.01       &    0.37      &  0.01       \\
 Lens Orientation             & \M 29\degpt1 &   0\degpt3  & \M 28\degpt5 &   0\degpt2  \\
 Source half-light            & 0\farcs054   & 0\farcs006  & 0\farcs045   & 0\farcs007  \\
 Source Orientation           &   0\degpt0   & Fixed       &   0\degpt0   & Fixed       \\
 Source Axis-Ratio            &    1.00      & Fixed       &    1.00      & Fixed       \\
\tableline
\end{tabular} 
\end{table} 
\end{center} 
%\clearpage
%HST 12531-2914 12:53:06.7 -29:14:31 21.8      urz00:0035  Cross         myung    Done Strong  Fixed  0.099 
\begin{figure}[H]
%\plotone{urz00g0035.eps}
\plotone{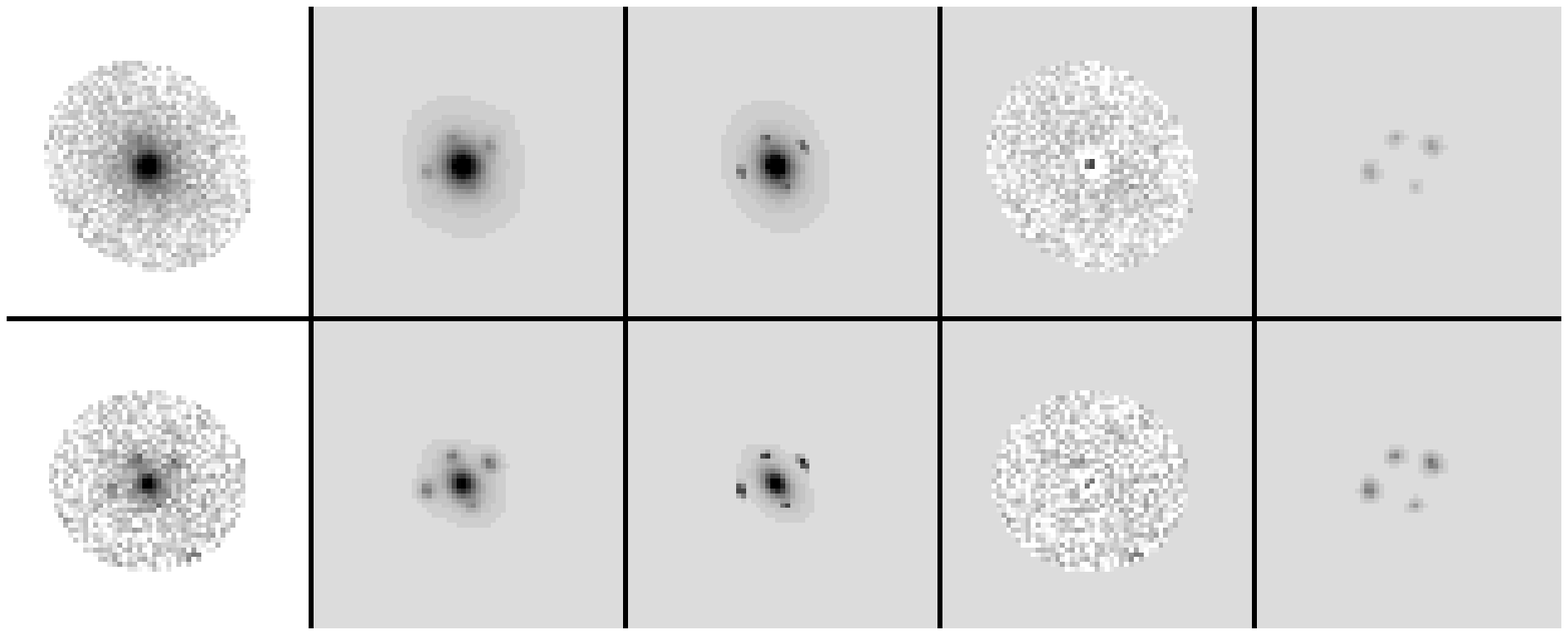}
\medskip
\caption{HST 12531-2914
The second HST lens candidate from Medium Deep Survey data
The lens:  F814W=21.8, F606W=23.8 ;
Each image of cross: F814W=26.8, F606W=26.7 }
\end{figure}

{\bf HST 12531-2914} is the second MDS HST ``Einstein Cross'' lens
candidate. The elliptical lensing galaxy has an apparent color
(V-I)=2.0 mag. The lensed source appears to be point-like with a color
(V-I)=0.2 mag. The impact radius is 0\farcs06. The image magnification
is about 2.3 magnitudes, making each component about 0.8 mag brighter
than the point-like source. The model gives a very good fit with a
normalized $\chi^2$ of unity.

%\clearpage
\begin{center}
\begin{table}[H]
\caption{}
\medskip
\begin{tabular}{ l c l c l }
\tableline
 Name           &  \multicolumn{4}{c}{HST12531\M 2914} \\
\tableline
\tableline
 Equatorial(J2000) & \multicolumn{1}{c}{12:53:06.7} & \multicolumn{1}{c}{\M 29:14:30} & \multicolumn{2}{c}{\VT=127\degpt77}\\
 HST WFPC2  &  \multicolumn{4}{c}{Griffiths GO-PAR:5369 15-Feb-1995} \\
 Dataset[g][x,y]& \multicolumn{4}{c}{U26K7G04T[3][755,326]}  \\ 
 MDS Field:id lens galaxy & \multicolumn{4}{c}{urz00:0035}\\
MLE%\begin{tabular}{l  r r c c c c  r r c c c c  }
Mag% Configurationm & X    & Y    &   V   & $\pm$ & V\M I & $\pm$ \\
% Elliptical     &  0   &  0   & 23.77 & 0.06  & 1.95  & 0.07  \\
%    A           & \M 7 & \M 3 & 27.02 & 0.15  & 0.25  & 0.28  \\
%    B           &  5   &  2   & 26.89 & 0.15  & 0.43  & 0.23  \\
%    C           & \M 3 &  3   & 26.72 & 0.11  & 0.34  & 0.21  \\
%    D           &  2   & \M 5 & 27.51 & 0.24  & 0.82  & 0.34  \\
\\
 HST WFPC2 Filter             &\multicolumn{2}{c}{F814W}   &\multicolumn{2}{c}{F606W}   \\
 Exposure seconds             &\multicolumn{2}{c}{4 x 2100}&\multicolumn{2}{c}{3 x 1800}\\
% log (integrated S/N Ratio)  &\multicolumn{2}{c}{2.69}    &\multicolumn{2}{c}{3.14}    \\
 Fitted Parameter             &  MLE      & $\pm$ rms   &  MLE         & $\pm$     \\
\\
% Sky Mag. ($arcsec^{2}$)     &  21.959      & 0.002     &  22.499      & 0.002       \\
 Total  of lens galaxy        &   21.76      &  0.02     &   23.81      &  0.04       \\
% X centroid 0\farcs1 Pix     &  754.68      &  0.01     &  754.64      &  0.03       \\
% Y centroid 0\farcs1 Pix     &  326.13      &  0.02     &  326.48      &  0.03       \\
 Half-light radius            & 0\farcs205   & 0\farcs007& 0\farcs180   & 0\farcs011  \\
 Position Angle light         &  18\degpt7   &   4\degpt0&  39\degpt0   &   6\degpt2  \\
 Axis Ratio of the light      &    0.78      &  0.03     &    0.64      &  0.07       \\
Mag Mag. of lensed source     &   27.42      &  0.13     &   27.65      &  0.14       \\
 Source X offset 0\farcs1 Pix &   \M 0.40    & Fixed     &   \M 0.40    &  0.06       \\
 Source Y offset 0\farcs1 Pix &    0.50      & Fixed     &    0.50      &  0.07       \\
 Critical Radius              & 0\farcs644   & Fixed     & 0\farcs644   & 0\farcs006  \\
 Axis Ratio Lens mass         &    0.44      & Fixed     &    0.44      &  0.04       \\
 Lens Orientation             &  22\degpt5   & Fixed     &  22\degpt5   &   0\degpt9  \\
 Source half-light            & 0\farcs030   & Fixed     & 0\farcs030   & 0\farcs009  \\
 Source Orientation           &   0\degpt0   & Fixed     &   0\degpt0   & Fixed       \\
 Source Axis-Ratio            &    1.00      & Fixed     &    1.00      & Fixed       \\
\tableline   
\end{tabular}
\end{table}
\end{center}
%\clearpage
%HST 14164+5215 14 16 25.2 +52 14 31 21.6      u26xi:0017  pair          kavan    Done Strong 0.092 0.087
\begin{figure}[H]
%\plotone{u26xig0017.eps}
\plotone{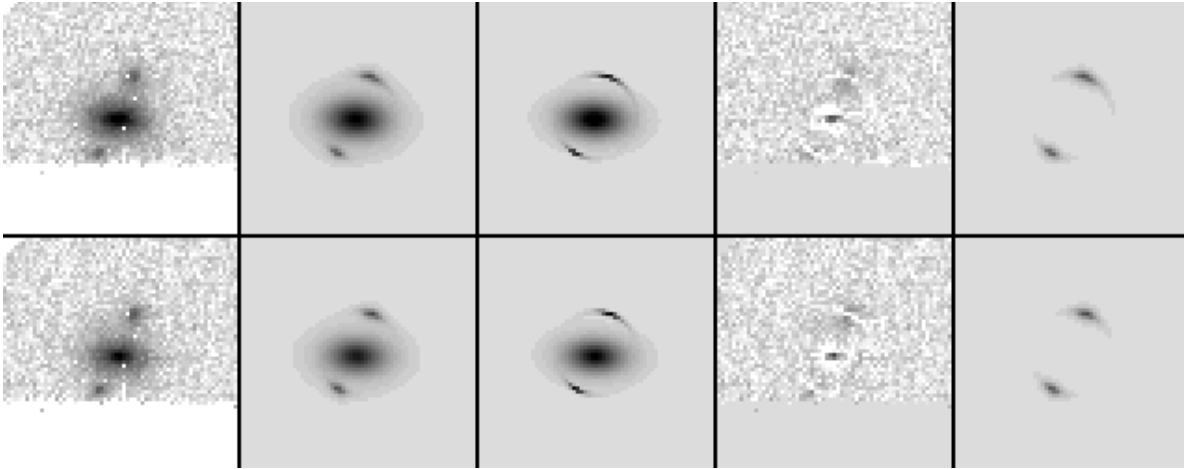}
\medskip
\caption{HST 14164+5215, A pair in the GWS. The model is tentative
since the galaxy was observed so near the WFPC2 pyramid edge and
differential image distortion is probably significant. The observed
lensed images are clearly not as stretched as the model.  The lens:
F814W=20.8, F606W=21.8 Each image of pair: F814W=23.6, F606W=24.3}
\end{figure}

{\bf HST 14164+5215} is a pair of images symmetrically place around a
brighter galaxy in GWS. It is a difficult candidate to model since the
differential image distortion so close to the WFPC2 pyramid edge is
unknown and probably significant. It was ignored in the preliminary
model presented here and may account for the poor fit if the candidate
is confirmed spectroscopically. Since the lensing galaxy is a disk
with a 20\% bulge, a disk profile was adopted for the single component
lens model. The bulge is apparent at the center in the residual
image. The lens galaxy has an apparent color (V-I)=1.0 mag. The lensed
source appears to be point-like with a color (V-I)=0.7 mag. The impact
radius is 0\farcs1. The image magnification is about 3.5 magnitudes,
making each component about 2.7 mag brighter than the source. The
model gives a very good fit with a normalized $\chi^2$ just larger
than unity (1.07).

%\clearpage
\begin{center}
\begin{table}[H]
\caption{}
\medskip
\begin{tabular}{ l c l c l }
 Name                         & \multicolumn{4}{c}{HST 14164+5215}                               \\
\tableline
\tableline
 Equatorial(J2000) & \multicolumn{1}{c}{14:16:25.2} & \multicolumn{1}{c}{+52:14:31} & \multicolumn{2}{c}{\VT=32\degpt72}\\
Pix HST 2  & \multicolumn{4}{c}{Groth :5090 17-Mar-1994}\\
 Dataset[g][x,y] & \multicolumn{4}{c}{U26X0I01T[4][122,064]}\\
  Field:id lens galaxy & \multicolumn{4}{c}{u26xi:0017}\\
\\
 HST WFPC2 Filter             &\multicolumn{2}{c}{F814W}   &\multicolumn{2}{c}{F606W}   \\
 Exposure seconds             &\multicolumn{2}{c}{4 x 1100}&\multicolumn{2}{c}{4 x  700}\\
% log (integrated S/N Ratio)  &\multicolumn{2}{c}{3.42}    &\multicolumn{2}{c}{3.32}    \\
 Fitted Parameter             &  MLE         & $\pm$ rms   &  MLE         & $\pm$ rms    \\
\\
% Sky Mag. ($arcsec^{2}$)     &  22.247      & 0.003     &  22.799      & 0.003        \\
 Total Mag of lens galaxy     &   20.78      &  0.01     &   21.80      &  0.02        \\
% X centroid 0\farcs1 Pix     &  122.43      &  0.04     &  122.67      &  0.05        \\
% Y centroid 0\farcs1 Pix     &   64.40      &  0.02     &   64.57      &  0.03        \\
 Half-light radius            & 0\farcs517   & 0\farcs006& 0\farcs534   & 0\farcs009   \\
 Orientation of light         &  87\degpt5   &   0\degpt8&  88\degpt2   &   1\degpt0   \\
 Axis Ratio of the light      &    0.57      &  0.01     &    0.56      &  0.01        \\
 . of lensed source           &   26.67      &  0.09     &   27.40      &  0.05        \\
 Source X offset 0\farcs1     &    0.43      &  0.02     &    0.40      &  0.02        \\
 Source Y offset 0\farcs1 Pix &    0.91      &  0.03     &    0.87      &  0.04        \\
 Critical Radius              & 1\farcs090   & 0\farcs003& 1\farcs099   & 0\farcs003   \\
 Axis Ratio Lens mass         &    0.96      &  0.01     &    0.98      &  0.01        \\
 Lens Orientation             & \M 45\degpt0 &   3\degpt4& \M 61\degpt5 &   5\degpt0   \\
 Source half-light            & 0\farcs013   & 0\farcs003& 0\farcs010   & 0\farcs001   \\
 Source Orientation           &   0\degpt0   & Fixed     &   0\degpt0   & Fixed        \\
 Source Axis-Ratio            &    1.00      & Fixed     &    1.00      & Fixed        \\
\tableline  
\end{tabular}
\end{table}
\end{center}  
%\clearpage
%HST 15433+5352 15 43 20.9 +53 51 52 21.3      uvd01:0014  Smiley        ejo      Done Strong 0.286 0.277 0.280
\begin{figure}[H]
%\plotone{uvd01g0014.eps}
\plotone{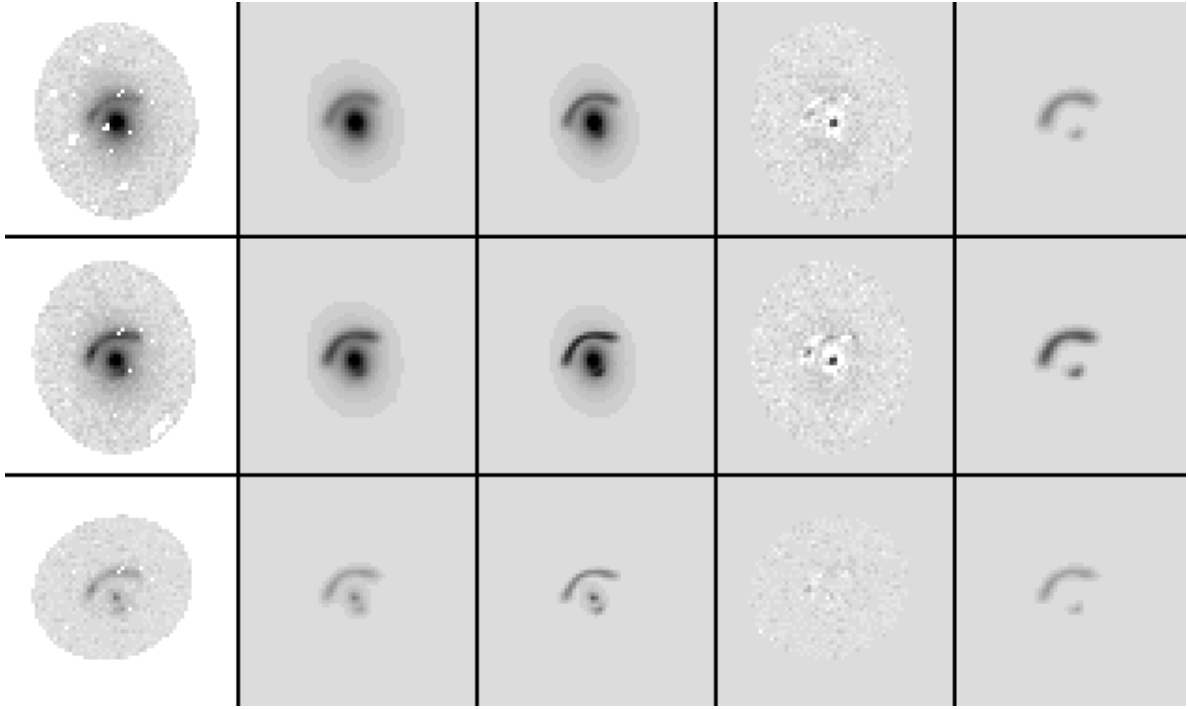}
\medskip
\caption{HST 15433+5352,
A Strong 3+1 lens of a blue object.
The lens: F814W=20.2, F606W=21.5, F450W=23.6
The Arc:  F814W=22.3, F606W=22.8, F450W=23.2  }
\end{figure}

{\bf HST 15433+5352} is very good strong lens candidate. The red
elliptical lensing galaxy has an apparent color (V-I)=1.4 mag, and
(B-V)=2.1 mag. The much bluer lensed source appears to be extended
with a color (V-I)=0.4 mag and (B-V)=0.4 mag. The impact radius is
0\farcs16. The image magnification is about 2.2 magnitudes. The model
gives a reasonably good fit with a normalized $\chi^2$ just larger
than unity.

%\clearpage
\begin{center}
\begin{table}[H]
\caption{}
\medskip
\begin{tabular}{l  c l c l c l }
\tableline
Name & \multicolumn{6}{c}{HST 15433+5352} \\
\tableline
\tableline
 Equatorial(J2000) & \multicolumn{2}{c}{15:43:20.9} & \multicolumn{2}{c}{+53:51:52} & \multicolumn{2}{c}{\VT=261\degpt77}\\
 HST WFPC2  & \multicolumn{6}{c}{GRIFFITHS GO-PAR:5971 03-Sep-1995}\\
 Dataset[g][x,y] & \multicolumn{6}{c}{U2OQ3F01T[2][068,715]}\\
 MDS Field:id lens galaxy & \multicolumn{6}{c}{uvd01:0014}\\
\\
  2 Filter             &\multicolumn{2}{c}{F814W}   &\multicolumn{2}{c}{F606W}   &\multicolumn{2}{c}{F450W}   \\
 Exposure seconds             &\multicolumn{2}{c}{2 x 3000}&\multicolumn{2}{c}{3 x 3000}&\multicolumn{2}{c}{3 x 2933}\\
% log (integrated S/N Ratio)  &\multicolumn{2}{c}{3.59}    &\multicolumn{2}{c}{3.61}    &\multicolumn{2}{c}{2.99}    \\
 Fitted Parameter             &  MLE         & $\pm$    &  MLE         & $\pm$ rms   &  MLE         & $\pm$ rms   \\
\\
% Sky Mag. ($arcsec^{2}$)     &  22.244      & 0.003      &  22.760      & 0.002     &  23.503      & 0.006       \\
 Total Mag of lens galaxy     &   20.20      &  0.01      &   21.60      &  0.01     &   23.67      &  0.09       \\
% X centroid 0\farcs1 Pix     &   68.44      &  0.01      &   68.24      &  0.01     &   68.32      &  0.03       \\
% Y centroid 0\farcs1 Pix     &  714.61      &  0.01      &  714.68      &  0.01     &  715.01      &  0.05       \\
 Half-light radius            & 0\farcs456   & 0\farcs009 & 0\farcs453   & 0\farcs002& 0\farcs385   & 0\farcs044  \\
 Orientation of light         &  10\degpt1   &   1\degpt3 &   9\degpt5   &   1\degpt8&  19\degpt3   &   7\degpt2  \\
 Axis Ratio of the light      &    0.75      &  0.01      &    0.78      &  0.01     &    0.67      &  0.08       \\
 Mag. of lensed source        &   24.51      &  0.04      &   24.95      &  0.02     &   25.39      &  0.06       \\
 Source X offset 0\farcs1 Pix &   \M 0.57    &  0.06      &   \M 0.42    &  0.03     &   \M 0.52    &  0.01       \\
 Source Y offset 0\farcs1     &    1.58      &  0.04      &    1.54      &  0.02     &    1.55      &  0.04       \\
 Critical Radius              & 0\farcs588   & 0\farcs006 & 0\farcs576   & 0\farcs002& 0\farcs584   & 0\farcs004  \\
 Axis Ratio Lens mass         &    0.50      &  0.02      &    0.56      &  0.01     &    0.55      &  0.01       \\
 Lens Orientation             &  29\degpt0   &   1\degpt0 &  31\degpt2   &   0\degpt5&  28\degpt3   &   0\degpt9  \\
 Source half-light            & 0\farcs147   & 0\farcs003 & 0\farcs146   & 0\farcs003& 0\farcs155   & 0\farcs007  \\
 Source Orientation           & \M 70\degpt2 &   2\degpt6 & \M 66\degpt6 &   0\degpt9& \M 69\degpt3 &   1\degpt6  \\
 Source Axis-Ratio            &    0.46      &  0.03      &    0.35      &  0.01     &    0.33      &  0.02       \\
\tableline 
\end{tabular}
\end{table}
\end{center}
%\clearpage
%HST 01247+0352 01 24 44.4 +03 52 00 23.9      uci10:0034  pair, Quad ?  ejo      Done Strong 0.742 0.730
\begin{figure}[H]
%\plotone{uci10g0034.eps}
\plotone{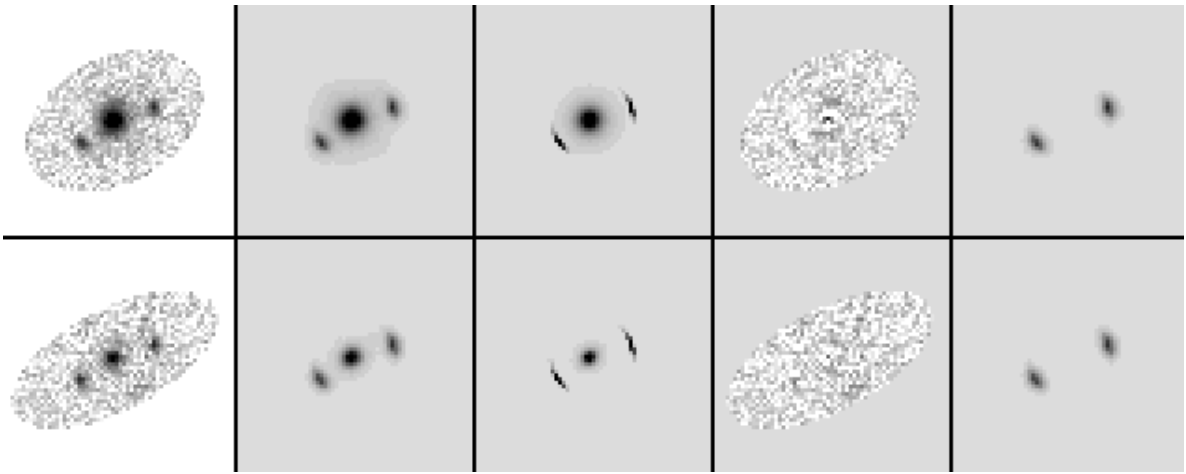}
\medskip
\caption{HST 01247+0352,
A pair (possible Quad) around an E0.
Fainter two images visible on screen display.
The lens:           F814W=21.9, F606W=24.0 ;
Each image of Pair: F814W=24.6, F606W=25.2  }
\end{figure}

{\bf HST 01247+0352} is a good candidate for a strong lensed pair. The
red spherical elliptical lensing galaxy (E0) has an apparent color
(V-I)=2.1 mag. The bluer lensed source appears to be point-like with a
color (V-I)=0.7 mag. Two much fainter images can be seen near the
detection limit which might make this a Quad system. The impact radius
is 0\farcs08. The image magnification is about 3.5 magnitudes, making
each component about 2.8 mag brighter than the source. The model gives
a reasonably good fit with a normalized $\chi^2$ close to unity.

%\clearpage
\begin{center}
\begin{table}[H]
\caption{}
\medskip
\begin{tabular}{l c l c l }
\tableline
 Name                           & \multicolumn{4}{c}{HST 01247+0352}                               \\
\tableline
\tableline
 Equatorial(J2000) & \multicolumn{1}{c}{01:24:44.4} & \multicolumn{1}{c}{+03:52:00} & \multicolumn{2}{c}{\VT=67\degpt18}\\
 HST WFPC2  & \multicolumn{4}{c}{GRIFFITHS GO-PAR:6251 01--1995}\\
 Dataset[g][x,y] & \multicolumn{4}{c}{U2OS3701T[3][548,294]}\\
  Field:id lens galaxy & \multicolumn{4}{c}{10:0034}\\
\\
 HST 2 Filter                 &\multicolumn{2}{c}{F814W}   &\multicolumn{2}{c}{F606W}   \\
 Exposure seconds             &\multicolumn{2}{c}{4 x 2700}&\multicolumn{2}{c}{3 x 1600}\\
% log (integrated S/N Ratio)  &\multicolumn{2}{c}{3.02}    &\multicolumn{2}{c}{2.56}    \\
 Fitted Parameter             &  MLE         & $\pm$ rms   &           & $\pm$ rms    \\
\\
% Sky Mag. ($arcsec^{2}$)     &  21.019      & 0.001      &  21.537      & 0.001      \\
 Total Mag of lens galaxy     &   21.93      &  0.02      &   24.01      &  0.06      \\
% X centroid 0\farcs1 Pix     &  547.95      &  0.01      &  547.90      &  0.02      \\
% Y centroid 0\farcs1 Pix     &  294.60      &  0.01      &  294.78      &  0.03      \\
 Half-light radius            & 0\farcs097   & 0\farcs005 & 0\farcs056   & 0\farcs011 \\
 Orientation of light         & \M 69\degpt4 &  39\degpt8 & \M 68\degpt4 &  53\degpt1 \\
 Axis Ratio of the light      &    0.97      &  0.05      &    0.74      &  0.18      \\
 Mag. of lensed source        &   27.27      &  0.04      &   27.93      &  0.05      \\
 Source X offset 0\farcs1 Pix &    0.72      &  0.01      &    0.72      &  0.02      \\
 Source Y offset 0\farcs1 Pix &    0.36      &  0.01      &    0.35      &  0.01      \\
 Critical Radius              & 1\farcs085   & 0\farcs002 & 1\farcs096   & 0\farcs002 \\
 Axis Ratio Lens mass         &    0.96      & Fixed      &    0.96      & Fixed      \\
 Lens Orientation             & \M 14\degpt6 &   2\degpt5 & \M 15\degpt6 &   3\degpt6 \\
 Source half-light            & 0\farcs010   & 0\farcs001 & 0\farcs010   & 0\farcs001 \\
 Source Orientation           &   0\degpt0   & Fixed      &   0\degpt0   & Fixed      \\
 Source Axis-Ratio            &    1.00      & Fixed      &    1.00      & Fixed      \\
\tableline
\end{tabular}
\end{table}
\end{center}
%\clearpage
%HST 01248+0351 01 24 45.6 +03 51 06 23.4      uci10:0050  pair disk     kristian Done Strong 0.125 0.114
\begin{figure}[H]
%\plotone{uci10g0050.eps}
\plotone{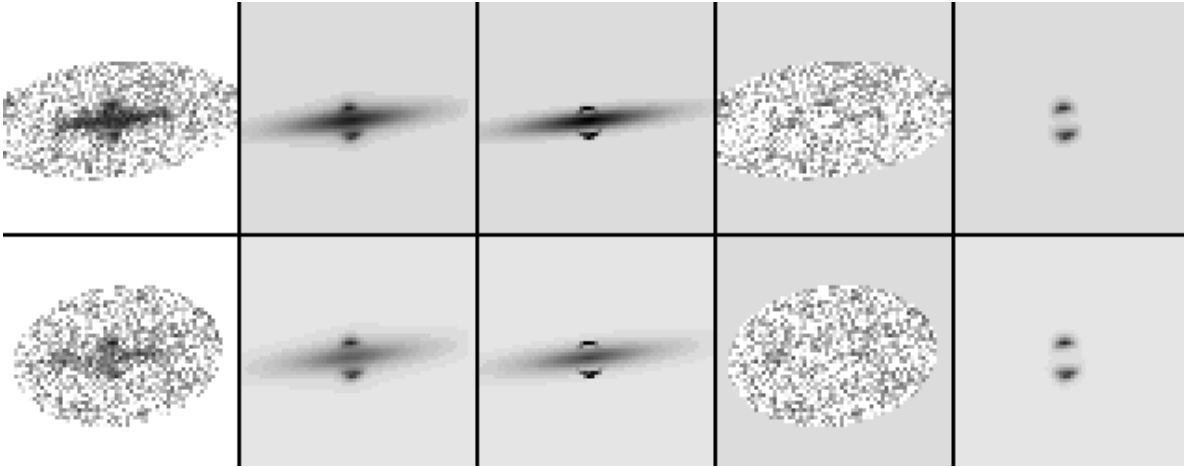}
\medskip
\caption{HST 01248+0351,
A possible lensed pair about minor axis of edge
on disk galaxy. The critical radius is 0\farcs4.
The lens:           F814W=22.6, F606W=23.8 ;
Each image of Pair: F814W=23.1, F606W=23.3  }
\end{figure}

{\bf HST 01248+0351} is a good candidate for a strong lensed pair in
the same WFPC2 field as HST 01247+0352. The edge-on disk lensing
galaxy has a color (V-I)=1.2 mag. The bluer lensed source appears to
be point-like with a color (V-I)=0.7 mag. The impact radius is
0\farcs05. The image magnification is about 3.1 magnitudes, making
each component about 2.4 mag brighter than the source. The model gives
a good fit with a normalized $\chi^2$ close to unity. Spectroscopy is
required to confirm this candidate as a gravitational lens. It is
included in this sample as the best candidate for lensing by a disk
(\cite{kee98}).

%\clearpage
\begin{center}
\begin{table}[H]
\caption{}
\medskip
\begin{tabular}{l c l c l }
\tableline
 Name                           & \multicolumn{4}{c}{HST 01248+0351}                               \\
\tableline
\tableline
 Equatorial(J2000) & \multicolumn{1}{c}{01:24:45.6} & \multicolumn{1}{c}{+03:51:06} & \multicolumn{2}{c}{\VT=67\degpt18}\\
 HST WFPC2  & \multicolumn{4}{c}{GRIFFITHS GO-PAR:6251 01-Jul-1995}\\
 Dataset[g][x,y] & \multicolumn{4}{c}{U2OS3701T[3][105,670]}\\
 MDS Field:id lens galaxy & \multicolumn{4}{c}{10:0050}\\
\\
  2 Filter             &\multicolumn{2}{c}{F814W}   &\multicolumn{2}{c}{F606W}   \\
 Exposure seconds             &\multicolumn{2}{c}{4 x 2700}&\multicolumn{2}{c}{3 x 1600}\\
% log (integrated S/N Ratio)  &\multicolumn{2}{c}{2.76}    &\multicolumn{2}{c}{2.27}    \\
 Fitted Parameter             &  MLE         & $\pm$    &  MLE         & $\pm$ rms    \\
\\
% Sky Mag. ($arcsec^{2}$)     &  20.996      & 0.001     &  21.516      & 0.001      \\
 Total Mag of lens galaxy     &   22.60      &  0.06     &   23.81      &  0.12      \\
% X centroid 0\farcs1 Pix     &  103.87      &  0.10     &  104.13      &  0.13      \\
% Y centroid 0\farcs1 Pix     &  669.64      &  0.04     &  670.19      &  0.04      \\
 Half-light radius            & 1\farcs247   & 0\farcs054& 1\farcs411   & 0\farcs035 \\
 Orientation of light         & \M 83\degpt7 &   0\degpt6& \M 83\degpt5 &   1\degpt4 \\
 Axis Ratio of the light      &    0.12      &  0.01     &    0.15      &  0.03      \\
 Mag. of lensed source        &   27.78      &  0.14     &   28.53      &  0.13      \\
 Source X offset 0\farcs1 Pix &    0.05      &  0.02     &    0.04      &  0.02      \\
 Source Y offset 0\farcs1 Pix &   \M 0.46    &  0.04     &   \M 0.48    &  0.04      \\
 Critical Radius              & 0\farcs369   & 0\farcs004& 0\farcs421   & 0\farcs004 \\
 Axis Ratio Lens mass         &    1.00      & Fixed     &    1.00      & Fixed      \\
 Lens Orientation             &   0\degpt0   & Fixed     &   0\degpt0   & Fixed      \\
 Source half-light            & 0\farcs011   & 0\farcs003& 0\farcs010   & 0\farcs003 \\
 Source Orientation           &   0\degpt0   & Fixed     &   0\degpt0   & Fixed      \\
 Source Axis-Ratio            &    1.00      & Fixed     &    1.00      & Fixed      \\
\tableline    
\end{tabular}
\end{table}
\end{center}
%\clearpage
%HST 16302+8230 16 30 12.9 +82 29 59 23.1      urg01:0042  underground   ejo      Done Strong 0.193  Fixed  Fixed
\begin{figure}[H]
%\plotone{urg01g0042.eps}
\plotone{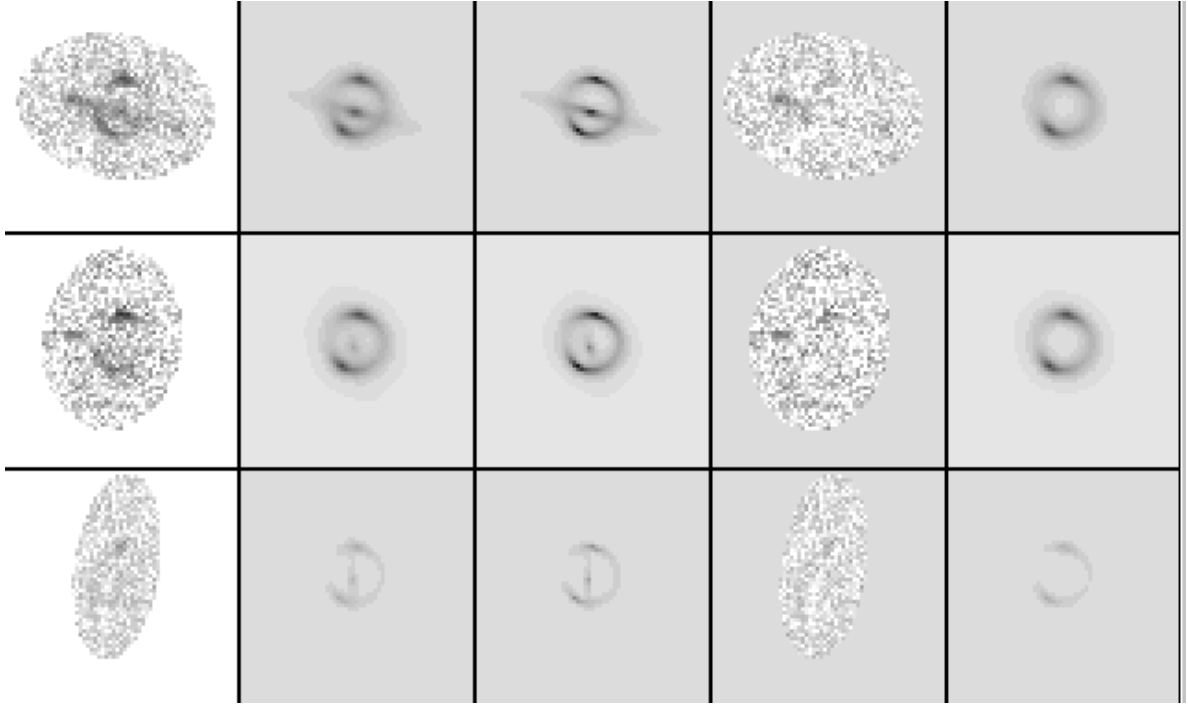}
\medskip
\caption{HST 16302+8230,
A possible Einstein ring centered on edge-on disk galaxy.
The lens: F814W=23.7, F606W=25.6, F450W=26.3 ;
The ring: F814W=22.8, F606W=23.7, F450W=25.0  }
\end{figure}

{\bf HST 16302+8230} could be an ``Einstein ring'' but clearly needs
spectroscopic confirmation, since it may also be a polar ring
galaxy. In our opinion it is the least probable of our top ten
candidate list but the most intriguing. It has been nicknamed the
``the London Underground'' since it resembles that logo.  The right
half of the the image is much fainter for an unknown reason. The
``ring'' is clearly much bluer than the central galaxy. The edge-on
disk lensing galaxy is barely visible in the V band and was not
detected in the B band which just shows a part of the ring.  If it is
a lensed source, then it appears to be extended with a color (V-I)=0.9
mag. The impact radius is 0\farcs14. The image magnification is about
2.1 magnitudes. The model gives a reasonably good fit with a
normalized $\chi^2$ close to unity.

%\clearpage
\begin{center}
\begin{table}[H]
\caption{}
\medskip
\begin{tabular}{l  c l c l cl  }
\tableline
 Name                           & \multicolumn{6}{c}{HST 16302+8230}                      \\
\tableline
\tableline
 Equatorial(J2000) & \multicolumn{2}{c}{16:30:12.9} & \multicolumn{2}{c}{+82:29:59} & \multicolumn{2}{c}{\VT=353\degpt22}\\
 HST WFPC2  & \multicolumn{6}{c}{GROTH  GTO-PAR:6254 06-Jun-1996}\\
 Dataset[g][x,y] & \multicolumn{6}{c}{U203T[2][110,590]}\\
 MDS Field:id lens galaxy & \multicolumn{6}{c}{01:0042}\\
\\
  HST WFPC2 Filter            &\multicolumn{2}{c}{F814W}   &\multicolumn{2}{c}{F606W}   &\multicolumn{2}{c}{F450W}   \\
 Exposure seconds             &\multicolumn{2}{c}{5 x 1100}&\multicolumn{2}{c}{3 x  900}&\multicolumn{2}{c}{7 x 1157}\\
% log (integrated S/N Ratio)  &\multicolumn{2}{c}{2.72}    &\multicolumn{2}{c}{2.40}    &\multicolumn{2}{c}{1.71}    \\
 Fitted Parameter             &  MLE         & $\pm$ rms   & MLE         & $\pm$ rms   &  MLE         & $\pm$ rms    \\
\\
% Sky Mag. ($arcsec^{2}$)     &  21.898      & 0.003      &  22.371      & 0.002      &  23.133      & 0.007 \\
 Total Mag of lens galaxy     &   23.72      &  0.13      &   25.70      &            &   Dropout    &   -   \\
% X centroid 0\farcs1 Pix     &  110.52      &  0.08      &  110.26      &  0.09      &  109.66      &  0.36 \\
% Y centroid 0\farcs1 Pix     &  591.75      &  0.05      &  591.81      &  0.06      &  592.69      &  0.18 \\
 Half-light radius            & 1\farcs398   & 0\farcs034 &     -        &   -        &     -        &   -   \\
 Orientation of light         &  75\degpt0   &   1\degpt6 &     -        &   -        &     -        &   -   \\
 Axis Ratio of the light      &    0.11      &  0.03      &     -        &   -        &    0.06      &  0.12 \\
 Mag. of lensed source        &   24.93      &  0.08      &   25.82      &  0.06      &   27.06      &  0.17 \\
 Source X offset 0\farcs1 Pix &    0.46      &  0.03      &    0.46      & Fixed      &    0.46      & Fixed \\
 Source Y offset 0\farcs1 Pix &    1.34      &  0.05      &    1.34      & Fixed      &    1.34      & Fixed \\
 Critical Radius              & 0\farcs734   & 0\farcs004 & 0\farcs734   & Fixed      & 0\farcs734   & Fixed \\
 Axis Ratio Lens mass         &    0.86      &  0.02      &    0.86      & Fixed      &    0.86      & Fixed \\
 Lens Orientation             & \M 79\degpt0 &   2\degpt9 & \M 79\degpt0 & Fixed      & \M 79\degpt0 & Fixed \\
 Source half-light            & 0\farcs290   & 0\farcs022 & 0\farcs290   & Fixed      & 0\farcs290   & Fixed \\
 Source Orientation           &  63\degpt0   &  18\degpt6 &  62\degpt2   & Fixed      &  62\degpt2   & Fixed \\
 Source Axis-Ratio            &    0.74      &  0.10      &    0.74      & Fixed      &    0.74      & Fixed \\
\tableline
\end{tabular}
\end{table}
\end{center}
%\clearpage
%HST 16309+8230 16 30 52.7 +82 30 12 21.2      urg01:0010  eye in sky    ejo      Done Arc 1.32  1.08  1.22
\begin{figure}[H]
%\plotone{urg01g0010.eps}
\plotone{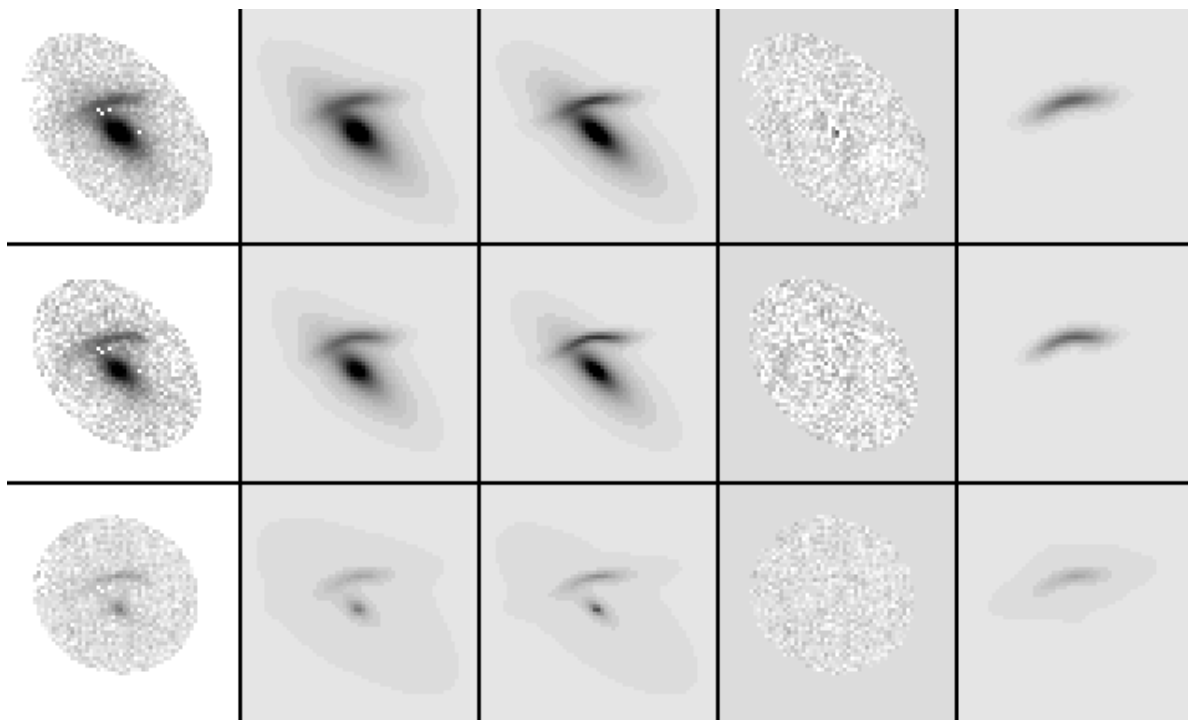}
\medskip
\caption{HST 16309+8230,
A galaxy lensed to an arc by an elliptical
The lens:          F814W=20.2, F606W=21.6, F450W=23.8 ;
The lensed galaxy: F814W=22.4, F606W=23.5, F450W=24.1  }
\end{figure}

{\bf HST 16309+8230} is an arc in the same field as HST 16302+8230.
The lensing elliptical galaxy has an apparent (V-I) color 1.4 mag and
(B-V)=2.2 mag. The significantly distorted lensed source is an
extended edge-on disk-like galaxy with color (V-I)=1.4 mag, and
(B-V)=0.4 mag.  The impact radius is 0\farcs5. The image magnification
is about 0.9 magnitudes. The model gives a good fit with a normalized
$\chi^2$ close to unity.

%\clearpage
\begin{center}
\begin{table}[H]
\caption{}
\medskip
\begin{tabular}{l  c l c l cl  }
\tableline
 Name                           & \multicolumn{6}{c}{HST 16309+8230}                      \\
\tableline
\tableline
 Equatorial(J2000) & \multicolumn{2}{c}{16:30:52.7} & \multicolumn{2}{c}{+82:30:12} & \multicolumn{2}{c}{\VT=353\degpt22}\\
 HST WFPC2  & \multicolumn{6}{c}{GROTH  GTO-PAR:6254 06-Jun-1996}\\
 Dataset[g][x,y] & \multicolumn{6}{c}{U2OVBT03T[3][061,575]}\\
 MDS Field:id lens galaxy & \multicolumn{6}{c}{urg01:0010}\\
\\
 HST WFPC2 Filter              &\multicolumn{2}{c}{F814W}   &\multicolumn{2}{c}{F606W}   &\multicolumn{2}{c}{F450W}   \\
 Exposure seconds              &\multicolumn{2}{c}{5 x 1100}&\multicolumn{2}{c}{3 x  900}&\multicolumn{2}{c}{7 x 1157}\\
% log (integrated S/N Ratio)   &\multicolumn{2}{c}{3.52}    &\multicolumn{2}{c}{3.20}    &\multicolumn{2}{c}{2.60}    \\
 Fitted Parameter              &  MLE         & $\pm$ rms   &  MLE         & $\pm$ rms   &  MLE         & $\pm$ rms    \\
\\
% Sky Mag. ($arcsec^{2}$)      &  21.912      & 0.003      &  22.385      & 0.003      &  23.107      & 0.005      \\
 Total Mag of lens galaxy      &  20.16       &  0.01      &   21.58      &  0.02      &   23.75      &  0.11      \\
% X centroid 0\farcs1 Pix      &   61.54      &  0.01      &   61.36      &  0.01      &   61.54      &  0.07      \\
% Y centroid 0\farcs1 Pix      &  573.76      &  0.01      &  574.05      &  0.01      &  574.18      &  0.06      \\
 Half-light radius             & 0\farcs619   & 0\farcs009 & 0\farcs746   & 0\farcs018 & 0\farcs656   & 0\farcs083 \\
 Orientation of light          &  48\degpt1   &   0\degpt4 &  47\degpt3   &   0\degpt7 &  51\degpt8   &   4\degpt2 \\
 Axis Ratio of the light       &    0.36      &  0.01      &    0.37      &  0.01      &    0.40      &  0.07      \\
 Mag. of lensed source         &   22.96      &  0.04      &   24.40      &  0.07      &   24.76      &  0.06      \\
 Source X offset 0\farcs1 Pix  &   \M 0.74    &  0.06      &   \M 0.34    &  0.07      &   \M 0.10    &  0.16      \\
 Source Y offset 0\farcs1 Pix  &    4.99      &  0.08      &    5.07      &  0.07      &    4.87      &  0.11      \\
 Critical Radius               & 0\farcs380   & 0\farcs006 & 0\farcs468   & 0\farcs004 & 0\farcs397   & 0\farcs012 \\
 Axis Ratio Lens mass          &    0.82      &  0.04      &    0.43      &  0.03      &    0.82      & Fixed      \\
 Lens Orientation              &  30\degpt3   & 114\degpt6 &   4\degpt9   &   1\degpt7 &  31\degpt5   & Fixed      \\
 Source half-light             & 0\farcs477   & 0\farcs021 & 0\farcs421   & 0\farcs032 & 0\farcs477   & Fixed      \\
 Source Orientation            & \M 75\degpt9 &   1\degpt0 & \M 75\degpt1 &   1\degpt9 & \M 75\degpt9 & Fixed      \\
 Source Axis-Ratio             &    0.24      &  0.02      &    0.21      &  0.03      &    0.24      & Fixed      \\
\tableline
\end{tabular}
\end{table}
\end{center}
%\clearpage
%HST 12368+6212 12 36 49.0 +62 12 22 22.7      uhdfk:0056  HDF                    Done Strong Fixed  0.743  Fixed
\begin{figure}[H]
%\plotone{uhdfkg0056.eps}
\plotone{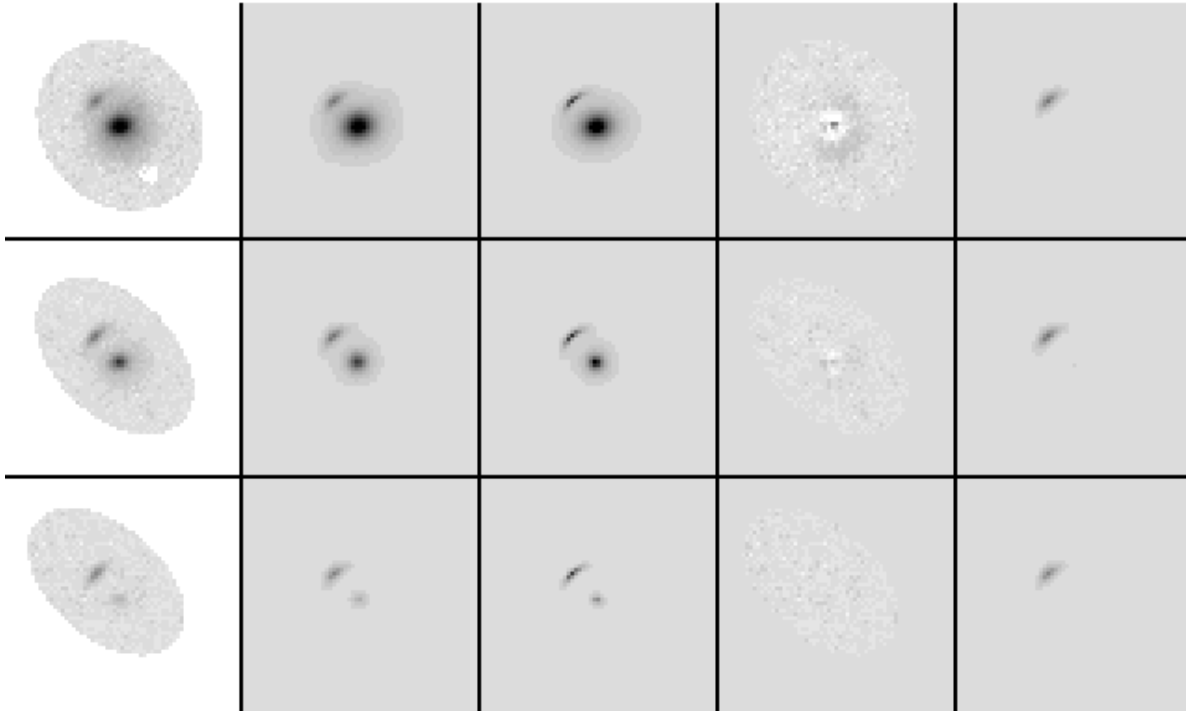}
\medskip
\caption{HST 12369+6212,
A blue galaxy strong-lensed to an arc by an elliptical galaxy in the Hubble Dee\
p Field.
The lens: F814W=22.0, F606W=24.0, F450W=26.1 ;
The Arc:  F814W=24.9, F606W=25.3, F450W=25.4 }
\end{figure}

{\bf HST 12368+6212} is an arc in the Hubble Deep Field (HDF). The
lensing elliptical galaxy has an apparent (V-I) color of 1.0 mag and
(B-V)=2.0 mag. The significantly distorted lensed source is extended
with a color (V-I)=0.3 mag, and (B-V)=0.2 mag. The impact radius is
0\farcs4. The image magnification is about 1.7 magnitudes. The model
gives a good fit with a normalized $\chi^2$ just larger than unity.

%\clearpage
\begin{center}
\begin{table}[H]
\caption{}
\medskip
\begin{tabular}{l  c l c l cl  }
\tableline
 Name                           & \multicolumn{6}{c}{HST 18078+4600}                      \\
\tableline
\tableline
 Equatorial(J2000) & \multicolumn{2}{c}{18:07:46.7} & \multicolumn{2}{c}{+45:59:56} & \multicolumn{2}{c}{\VT= 56\degpt36}\\
 HST WFPC2  & \multicolumn{6}{c}{GROTH  GTO-PAR:5092 27-Apr-1995}\\
 Dataset[g][x,y] & \multicolumn{6}{c}{U2807302T[3][364,506]}\\
 MDS Field:id lens galaxy & \multicolumn{6}{c}{uqc00:0029}\\
\\
 HST WFPC2 Filter             &\multicolumn{2}{c}{F814W}   &\multicolumn{2}{c}{F606W}   &\multicolumn{2}{c}{F450W}   \\
 Exposure seconds             &\multicolumn{2}{c}{2 x 2100}&\multicolumn{2}{c}{2 x  900}&\multicolumn{2}{c}{4 x 1875}\\
% log (integrated S/N Ratio)  &\multicolumn{2}{c}{3.28}    &\multicolumn{2}{c}{3.97}   &\multicolumn{2}{c}{2.55}    \\
 Fitted Parameter             &  MLE         & $\pm$ rms   &  MLE         & $\pm$ rms   &  MLE         & $\pm$ rms    \\
\\
% Sky Mag. ($arcsec^{2}$)     &  22.344      & 0.004      &  22.597      & 0.004      &  23.496      & 0.006      \\
  Total Mag of lens galaxy    &   20.66      &  0.02      &   22.60      &  0.09      &   Dropout    &  -         \\
% X centroid 0\farcs1 Pix     &  506.04      &  0.02      &  506.07      &  0.07      &  506.        &  -         \\
% Y centroid 0\farcs1 Pix     &  577.79      &  0.02      &  577.96      &  0.08      &  578.        &  -         \\
 Half-light radius            & 1\farcs480   & 0\farcs036 & 1\farcs502   & 0\farcs122 &     -        &  -         \\
 Orientation of light         &   0\degpt9   &   1\degpt0 &   0\degpt6   &   2\degpt9 &     -        &  -         \\
 Axis Ratio of the light      &    0.47      &  0.01      &    0.36      &  0.04      &     -        &  -         \\
 Mag. of lensed source        &   24.05      &  0.05      &   24.39      &  0.05      &   24.57      &  0.02      \\
 Source X offset 0\farcs1 Pix &    7.82      &  0.13      &    7.41      &  0.14      &    7.41      & Fixed      \\
 Source Y offset 0\farcs1 Pix &    3.82      &  0.04      &    4.08      &  0.10      &    4.09      & Fixed      \\
 Critical Radius              & 0\farcs456   & 0\farcs005 & 0\farcs449   & 0\farcs009 & 0\farcs449   & Fixed      \\
 Axis Ratio Lens mass         &    0.18      &  0.01      &    0.23      &  0.03      &    0.23      & Fixed      \\
 Lens Orientation             & \M 74\degpt7 &   0\degpt4 & \M 71\degpt9 &   1\degpt0 & \M 71\degpt7 & Fixed      \\
 Source half-light            & 0\farcs192   & 0\farcs010 & 0\farcs199   & 0\farcs008 & 0\farcs199   & Fixed      \\
 Source Orientation           &  64\degpt5   &   1\degpt6 &  64\degpt5   & Fixed      &  64\degpt5   & Fixed      \\
 Source Axis-Ratio            &    0.20      & Fixed      &    0.20      & Fixed      &    0.20      & Fixed      \\
\tableline
\end{tabular}
\end{table}
\end{center}
%\clearpage
%HST 18078+4600 18 07 46.7 +45 59 56 22.8      uqc00:0029  Arc in group  lwn      Done Arc 1.908  1.884  Fixed
% uqc00:0029 the lens with double nucleus
% uqc00:0086 the lensed image D+B(0.68) B B   
% uqc00:0033 the galaxy below D+B(0.49) D G
% uqc00:0079 the galaxy above   B       B G
\begin{figure}[H]
%\plotone{uqc00g0029.eps}
\plotone{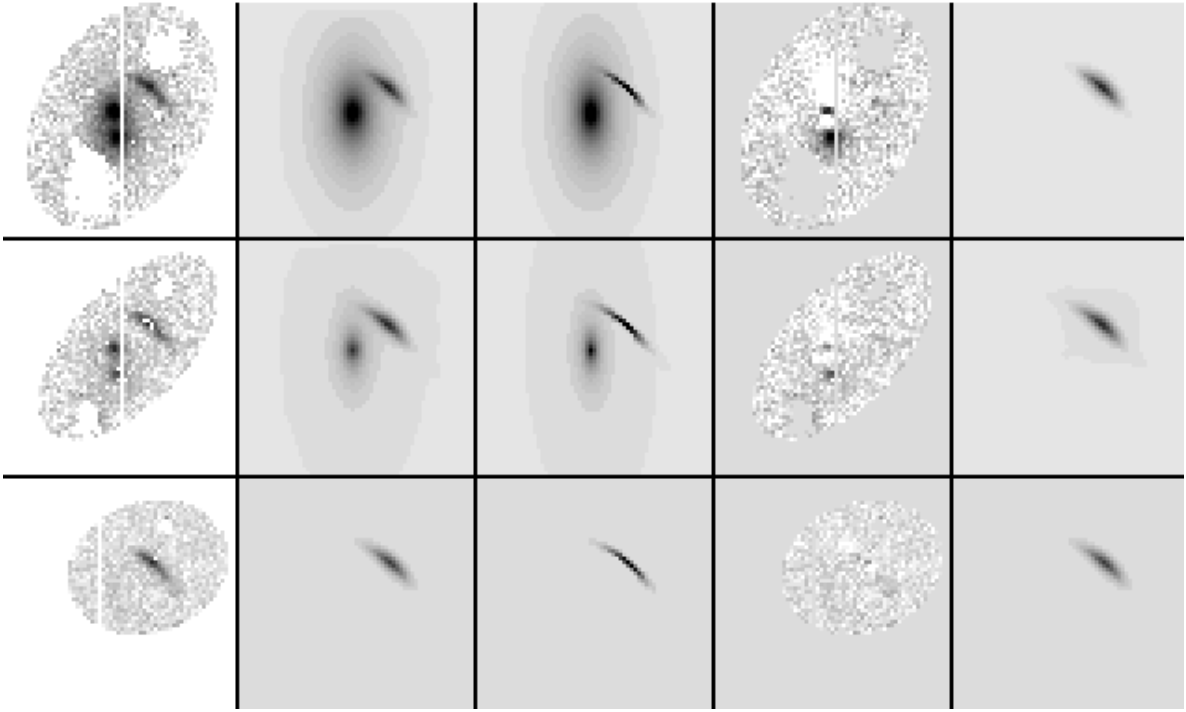}
\medskip
\caption{HST 18078+4600,
A blue galaxy lensed to an arc by a small cluster of galaxies.
The lens: F814W=20.7, F606W=22.7, F450W=dropout ;
The Arc:  F814W=23.5, F606W=23.7, F450W=23.9 }
\end{figure}

{\bf HST 18078+4600} is an arc caused by the gravitational potential
of a small group of 4 galaxies.  Since the software at hand, developed
for \cite{rat95} assumed a single lensing galaxy, the outer two galaxy
images were masked out in the image analysis which was used to derive
the approximate lens model.  The two central galaxies, with a color
(V-I)=2.0 mag, were merged in the model and were not detected in the B
band. The significantly distorted lensed extended source with colors
(V-I)=0.3 mag, and (B-V)=0.2 mag. The image magnification is about 0.7
magnitudes. The model gives a good fit in V and B with a normalized
$\chi^2$ just larger than unity, but a poor fit in I owing to the use
of a single lens galaxy approximation in the model.

%\clearpage
\begin{center}
\begin{table}[H]
\caption{}
\medskip
\begin{tabular}{l  c l c l c l  }
\tableline
 Name                         & \multicolumn{6}{c}{HST 12368+6212}                      \\
\tableline
\tableline
 Equatorial(J2000) & \multicolumn{2}{c}{12:36:49.0} & \multicolumn{2}{c}{+62:12:22} & \multicolumn{2}{c}{\VT=112\degpt00}\\
 HST WFPC2  & \multicolumn{6}{c}{WILLIAMS  DD:6337 27-Dec-1995}\\
 Dataset[g][x,y] & \multicolumn{6}{c}{U31P031AT[3][448,691]}\\
 MDS Field:id lens galaxy& \multicolumn{6}{c}{uhdfk:0056}\\
\\
%---------------------------- &\multicolumn{8}{c}{------------------------------------------------------------------}\\
 HST WFPC2 Filter             &\multicolumn{2}{c}{F814W}   &\multicolumn{2}{c}{F606W}   &\multicolumn{2}{c}{F450W}   \\
 Exposure seconds             &\multicolumn{2}{c}{ 58x2137}&\multicolumn{2}{c}{103x1050}&\multicolumn{2}{c}{ 58x2182}\\
% log (integrated S/N Ratio)   &\multicolumn{2}{c}{3.66}    &\multicolumn{2}{c}{3.31}    &\multicolumn{2}{c}{2.64}    \\
 Fitted Parameter             &  MLE         & $\pm$ rms   &  MLE         & $\pm$ rms   &  MLE          & $\pm$ rms   \\
\\
% Sky Mag. ($arcsec^{2}$)     &  22.116      & 0.001      &  22.754      & 0.001      &  23.268      & 0.001      \\
 Total Mag of lens galaxy     &   22.01      & Fixed      &   23.98      &  0.01      &   26.02      &  0.16      \\
% X centroid 0\farcs1 Pix     &  448.55      &  0.01      &  448.28      &  0.01      &  448.56      &  0.03      \\
% Y centroid 0\farcs1 Pix     &  690.20      &  0.01      &  690.74      &  0.01      &  690.96      &  0.03      \\
 Half-light radius            & 0\farcs283   & 0\farcs001 & 0\farcs316   & 0\farcs005 & 0\farcs332   & 0\farcs076 \\
 Orientation of light         & \M 77\degpt5 &   1\degpt5 & \M 81\degpt2 &   2\degpt2 &  \M 3\degpt2 & 114\degpt6 \\
 Axis Ratio of the light      &    0.82      &  0.01      &    0.71      &  0.02      &    1.00      &  0.03      \\
 Mag. of lensed source        &   26.66      &  0.02      &   27.00      &  0.04      &   27.17      &  0.02      \\
 Source X offset 0\farcs1 Pix &   \M 2.62    & Fixed      &   \M 2.62    &  0.02      &   \M 2.62    & Fixed      \\
 Source Y offset 0\farcs1 Pix &    3.45      & Fixed      &    3.45      &  0.03      &    3.45      & Fixed      \\
 Critical Radius              & 0\farcs583   & Fixed      & 0\farcs583   & 0\farcs004 & 0\farcs583   & Fixed      \\
 Axis Ratio Lens mass         &    0.58      & Fixed      &    0.58      &  0.01      &    0.58      & Fixed      \\
 Lens Orientation             &  30\degpt9   & Fixed      &  30\degpt9   &   0\degpt7 &  30\degpt9   & Fixed     \\
 Source half-light            & 0\farcs053   & Fixed      & 0\farcs053   & 0\farcs002 & 0\farcs053   & Fixed     \\
 Source Orientation           &   0\degpt0   & Fixed      &   0\degpt0   & Fixed      &   0\degpt0   & Fixed     \\
 Source Axis-Ratio            &    1.00      & Fixed      &    1.00      & Fixed      &    1.00      & Fixed      \\
\tableline
\end{tabular}
\end{table}
\end{center}
%\clearpage
\section{Survey Area}

The number of strong gravitational lenses as a function of redshift is
very sensitive to the value of the cosmological constant $\Lambda$
(\cite{fuk92,koc92}) and could be used as an estimator of the latter
quantity if all of the complex detection thresholds which govern
completeness are well understood and known quantitatively.

These ten gravitational lenses were discovered serendipitously on
WFPC2 images processed via the data analysis pipeline of the HST
Medium Deep Survey and were not found as a result of a very systematic
or quantitative search. In view of the very large range of data
quality amongst the pure parallel MDS observations and because of
overlapping fields, it is not possible to be precise about the number
of fields surveyed and we have deliberately rounded the numbers to
reflect the nature of approximate estimates. In the full survey we
processed over 500 WFPC2 fields, each covering about 4.77 square
arcmin of sky. Overlap between WFPC2 fields is on average about 20\%
for the MDS survey. All of the Gravitational lenses were however
discovered on about 130 WFPC2 fields which were designated ``priority
one'' for which we had three or more exposures in each of two or more
filters. Although only just over 10\% of the ``priority one'' fields
had exposures in the three BVI filters, 50\% of the lenses discovered,
have observations in all three filters. Since the pointings were
random, this implies a very strong selection effect in terms of data
quality.  Clearly we found the ``Top ten list'' of lenses in the best
data available to the Medium Deep Survey.

\section{Incompleteness}

Two factors distinguish these WFPC2 fields from the rest. More
exposures usually indicated a longer total exposure time and maybe
more importantly better cosmic ray rejection. Removal of cosmic rays
in single exposures clearly runs the risk of losing the lensed images
in the cleaning out of many cosmic rays, and this remains true for
stacks with two images which require cleaning out of the
non-negligible number of pixels hit by cosmic rays in both images. In
Figure~11 we show the limiting magnitude for completeness of the
object catalog as a function of total integration time, for all WFPC2
images processed through the MDS pipeline in the 3 main filters. The
``priority one'' MDS fields are plotted with circles, with filled
circles showing those fields on which gravitational lenses were
discovered.

\begin{figure}
\epsscale{.75}
%\plotfiddle{figlimag.eps}{6.0in}{-90.}{75.}{75.}{-288}{432}
\plotfiddle{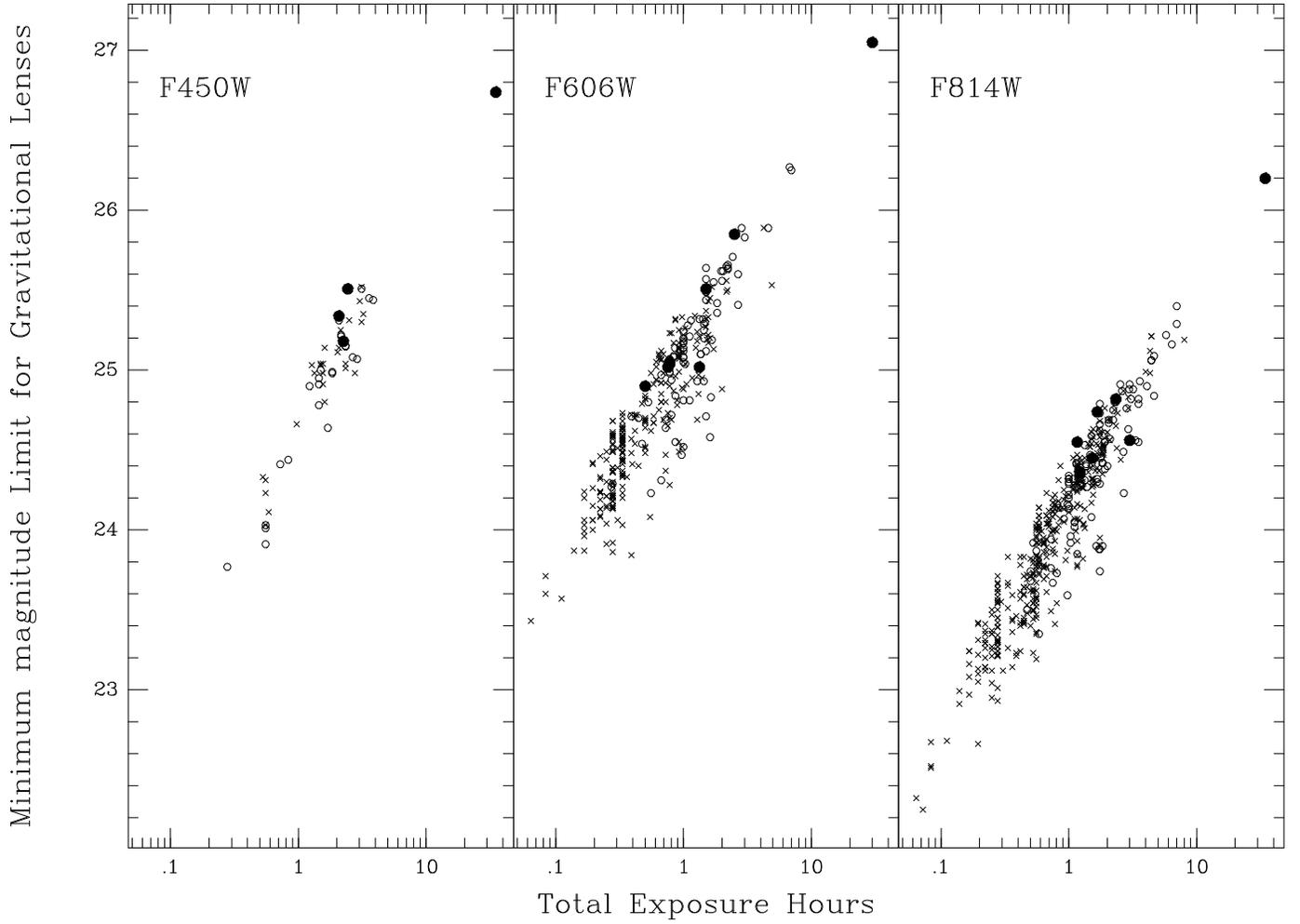}{6.0in}{-90.}{75.}{75.}{-288}{432}
\medskip
\caption{Limiting magnitude for completeness of WFPC2 processed
through the MDS pipeline as a function of total integration time.  The
priority one MDS fields are plotted with circles with the fields on
which gravitational lenses were discovered being filled. Illustrates
selection effects of lens discovery. }
\end{figure}

It appears that a limiting magnitude fainter than 25.1 mag in F450W,
24.8 in F606W and 24.3 in F814W, requiring total exposure times
typically longer than 1 hour, is needed for discovery of a
gravitational lens. Assuming that limiting magnitude is the only
factor, the survey of about 160 WFPC2 fields, less about 20\% overlap,
leads to an estimate of the total survey area of about 0.17 square
degrees.

The actual process of discovery of the lenses included a manual
component whereby the most likely candidates of gravitational lensing
were picked out during an inspection stage of the MDS pipeline or from
the residual images after removal of the maximum likelihood scale-free
model for the galaxy. Typically the lensed object was much bluer than
the lensing galaxy, and is seen better on the F450W or F606W filter
rather than F814W which was by default the first filter used for
parallel observations of limited total time on each field. The
availability of an image in other filters was useful to ensure that
the colors of the lensed images were about the same. So the fact that
all the lenses discovered in the MDS are on ``priority one'' fields is
probably not coincidental. If good images in two or more filters (MDS
priority one) was the only factor, then the survey consisted of about
130 WFPC2 fields with an estimated area of about 0.14 square
degrees. If we take both conditions i.e. limiting magnitude and MDS
``priority one'', then we have about 90 fields or about 0.1 square
degrees, which is a lower estimate.

An incomplete sample of gravitational lenses can, however, be used to
measure $\Lambda$ if we know or can average over unknown properties of
the lensing galaxies such as the mass distribution (\cite{ime97}).
The expected bias caused by the orientation of mass should be random
and could be averaged out. However a selection effect that is caused
by internal extinction within the lens galaxy may lead to a systematic
bias that does not average out. When a statistically significant
sample of spectroscopically confirmed candidates becomes available,
the MDS database is an ideal place for the comparison of the
properties of the lensing ellipticals and non-lensing ellipticals to
see if lenses are found preferentially in low extinction systems.

We have presented a ``Top Ten list'' of the lens candidates. In most
of these cases the lensed images are well resolved from the lensing
galaxy. Most of the candidates were picked while processing the data
through the MDS pipeline.
% Two high school students have independently looked at the residuals and 
% re-discovered the same candidates, although rejection was probably based 
% on our own personal bias passed on to them in their training process.
The current sample unfortunately suffers the same selection biases as
the serendipitously discovered sample of gravitational lenses known at
brighter magnitudes. At the expense of having a slightly incomplete
sample, the lenses modeled in this paper are selected to have the
highest probability of being good gravitational lens candidates.

If one lowers the standard of acceptance, one can pick out hundreds
more possible lens candidates, a small fraction of which may turn out
to be gravitational lenses. After removal of the smooth axis-symmetric
galaxy model images, the residuals sometimes configure symmetrically
into possible gravitational lens candidates. Most of these residual
images are probably bright regions of star formation within the very
faint spiral arms around the central bulge of a galaxy.  It is thus
very unlikely that most of these objects are lenses. A spectroscopic
study of a significant sample of these objects is unfortunately not
practical using current ground based observations, and these
candidates therefore do not represent a scientifically usable
sample. They have therefore been excluded from the present list.

\begin{figure}[H]
\epsscale{.4}
%\plotone{u2c45b0020.eps}
\plotone{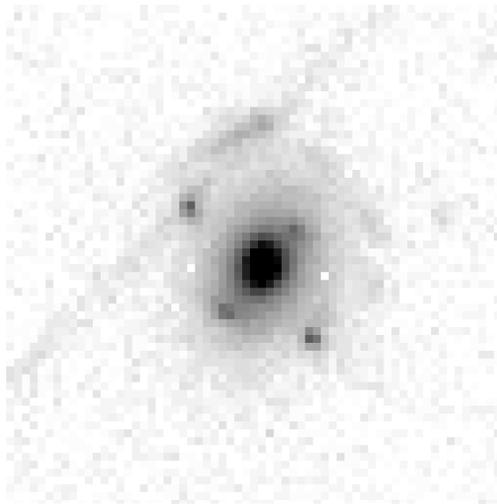}
\medskip
\caption[u2c45b0020.eps]{Spiral Galaxy HST 14113+5211 in MDS field
u2c45. The streak across is a diffraction spike from a bright star. }
\end{figure}

A case such as this was recently published by \cite{fsb98}. In fig~12
we show the region around this galaxy which appears to show an
extended face-on spiral arm structure with the bright regions aligned
along them.  In addition, the best-fit lens model of the observed
image configuration requires the adoption of a complex potential which
is significantly misaligned from the light distribution.  Although it
will remain a candidate until spectroscopic observation is done, we
suspect that these images are of star-formation regions, as found in
the case reported in \cite{gleag94}.

\section{Conclusions}

Gravitational lenses discovered using HST are especially useful
because the lens galaxies are at great distances (typically
$z_L\sim0.6$), allowing for an independent and new method for the
study of both galaxy evolution and cosmology.
 
A total of ten good candidates of gravitational lensing have been
discovered in the HST Medium Deep Survey (MDS) and archival primary
observations using WFPC2. Seven are ``strong lens'' candidates which
appear to have multiple images of the source. Three are cases where
the image of the source galaxy has been significantly distorted into
an arc. We have summarized the data on all ten candidates and
described them with simple models assuming singular isothermal
potentials.

\acknowledgments

This paper is based on observations with the NASA/ESA {\it Hubble
Space Telescope}, obtained at the Space Telescope Science Institute,
which is operated by the Association of Universities for Research in
Astronomy, Inc., under NASA contract NAS5-26555. The Medium-Deep
Survey was funded by STScI grant G02684 {\it et seq.} and by the HST
WFPC2 Science Team under JPL subcontract 960772 from NASA contract
NAS7-918.  We acknowledge contributions of Dr Lyman Neuschaefer who
was associated with the MDS pipeline.  We thank Kristin Wintersteen
and Julie Snyder who did High School Projects at CMU with the MDS
database. We thank Drs. Stefano Casertano, Myungshin Im and Joseph
Lehar for useful discussions, and our referee for useful comments.

\end{document}